\documentclass[%
 reprint,
superscriptaddress,
 amsmath,amssymb,
 aps,
prx,
]{revtex4-2}

\usepackage{graphicx}
\usepackage{dcolumn}
\usepackage{bm}
\RequirePackage[utf8]{inputenc}
\usepackage{xcolor}
\usepackage{siunitx}
 
\begin{document}

\preprint{APS/123-QED}

\title{Fresnel drag in space-time modulated metamaterials}

\author{P.A. Huidobro}%
\affiliation{Instituto de Telecomunicações, Insituto Superior T\'ecnico, University of Lisbon, Avenida Rovisco Pais 1,1049-001 Lisboa, Portugal}

\author{E. Galiffi}
\affiliation{Condensed Matter Theory Group, The Blackett Laboratory, Imperial College, SW7 2AZ, London, UK}
\author{S. Guenneau}
\affiliation{Department of Mathematics, Imperial College London, London SW7 2AZ, UK}
\affiliation{Aix Marseille Univ, CNRS, Centrale Marseille, Institut Fresnel, Marseille, France}
\author{R.V. Craster}
\affiliation{Department of Mathematics, Imperial College London, London SW7 2AZ, UK}

\author{J.B. Pendry}
\affiliation{Condensed Matter Theory Group, The Blackett Laboratory, Imperial College, SW72AZ, London, UK}




\date{\today}

\begin{abstract}
    A moving medium drags light along with it as measured by Fizeau and explained by Einstein's theory of special relativity. Here we show that the same effect can be obtained in a situation where there is no physical motion of the medium. Modulations of both the permittivity and permeability, phased in space and time in the form of travelling waves, are the basis of our model. Space-time metamaterials are represented by effective bianisotropic parameters, which can in turn be mapped to a moving homogeneous medium. Hence these metamaterials mimic a relativistic effect without the need for any actual material motion. We discuss how both the permittivity and permeability need to be modulated in order to achieve these effects, and we present an equivalent transmission line model.
\end{abstract}

\maketitle


\section{Introduction}

In 1818 Fresnel produced the aether drag hypothesis: a moving fluid appears to `drag' light along so that light travelling in opposite directions to the fluid flow would have different velocities \cite{Fresnel1818}. His extra velocity was related to but not equal to the velocity of the fluid. Although Fresnel's derivation was flawed, Fizeau in 1851 measured the drag effect and confirmed Fresnel's formula \cite{Fizeau}. A correct explanation followed in the wake of Einstein's theory of relativity, of which the aether drag experiment is one of the corner stones \cite{Einstein1905}.

\begin{figure}[t!]
    \begin{center}
        \includegraphics[width=\columnwidth]{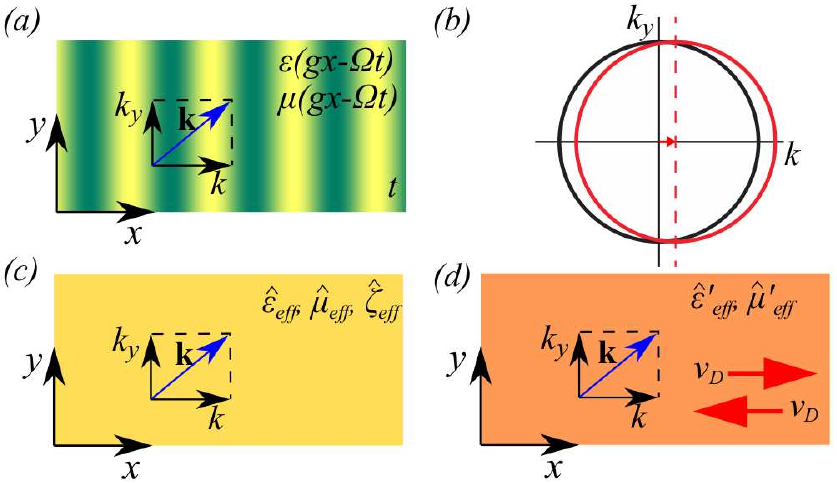}
        \caption{ Fresnel drag in space-time modulated metamaterials. (a) A wave propagates in a medium with space-time  travelling-wave modulated permittivity and permeability. The wave-vectors $k$ and $k_y$ are along the direction of the space-time grating and orthogonal to it, respectively. (b) Sketch of the low frequency dispersion surface in the $(k,k_y)$ plane. The isofrequency contours are ellipses centered at a finite value of $k$, reflecting a Fresnel drag effect along the modulation direction. (c) In the long wavelength limit the space-time modulated medium is mapped into a magnetoelectric medium with anisotropic effective permittivity $\hat\varepsilon_{\text{eff}}$, permeability $\hat\mu_{\text{eff}}$ and magnetoelectric coupling $\hat\zeta_{\text{eff}}$ parameters. (d) Equivalent moving medium interpretation: uniaxial medium with anisotropic effective permittivity $ \hat\varepsilon'_{\text{eff}}$ and permeability $ \hat\mu'_{\text{eff}}$ medium moving with velocity $v_D$.  }
        \label{fig:schematics}
    \end{center}
\end{figure}

While it may seem that physical movement of the fluid is an essential part of aether drag \cite{PhysRevApplied.10.047001,caloz2019spacetime}, here we come to the surprising conclusion that a time dependent system involving no physical movement of a medium can also produce a drag effect. We calculate the shifted dispersion surface, and show that the system can be represented as a bianisotropic metamaterial (see Fig. \ref{fig:schematics}), whose magnetoelectric coupling vanishes when either the dielectric or the magnetic modulation is switched off. Finally we propose a simple experiment which would demonstrate our results.

We consider a metamaterial whose permittivity and permeability are modulated in space and time following a travelling-wave form,
\begin{eqnarray}
    \epsilon(x,t) = \epsilon_m[1 + 2\alpha_e\cos(gx-\Omega t)] \\
    \mu(x,t)      = \mu_m[1 + 2\alpha_m\cos(gx-\Omega t)]
\end{eqnarray}
where $g$ and $\Omega$ are the spatial and temporal frequencies, $\alpha_{e,m}$ are the electric and magnetic modulation strengths, and $\epsilon_m$ and $\mu_m$ are the background permittivity and permeability of the medium. The profile moves with a phase velocity of $c_g=\Omega/g$ but we emphasize that the medium itself does not move: as a consequence the phase velocity can take any value between zero and infinity without violating special relativity. In addition, we note that we limit ourselves to situations where both $\epsilon$ and $\mu$ are greater than 1 so that a dispersionless approximation holds.  We shall concern ourselves with low-frequency, long-wavelength excitations, enabling the medium to be represented as a metamaterial with effective medium parameters which we calculate. Importantly, the drift velocity appearing in space-time metamaterials differs from both the modulation phase velocity and the conventional Fresnel-Einstein result, and it can even oppose the modulation phase velocity, when this is higher than the speed of light $c_m$ in the unmodulated medium.

Modulation of the electric permittivity in space and time has attracted much attention, as it gives rise to a plethora of exotic effects ranging from frequency-momentum transitions \cite{interbandWinn1999,lira2012electrically} to compression and amplification of electromagnetic signals \cite{cassedy1963dispersion,cassedy1967dispersion,PhysRevE.75.046607,sounas2017non,morgado2017negative,koutserimpas2018parametric,galiffibroadbandnonreciprocal2019} and even non-Hermitian and topological phenomena \cite{koutserimpas2018nonreciprocal,regensburger2012parity,lin2016photonic,fleury2016floquet,he2019floquet}. The directionality of space-time modulations such as travelling-wave modulations breaks time-reversal symmetry, which is reflected in non-reciprocal band diagrams \cite{PhysRevApplied.10.047001}. The breaking of reciprocity has been exploited in the realization of photonic isolators and circulators without the need of external static magnetic biasing \cite{yu2009complete,sounas2013giant,estep2014magnetic}. 
Recently, it was shown that in-phase modulations of the permittivity and the permeability with the same strength results in the closing of the high-frequency band gaps, at the same time as keeping the non-reciprocal character of these systems \cite{Taravati2018}. Here we further show that this non-reciprocity is in fact accompanied by a Fresnel drag and we present an effective medium model that illuminates its origin. 

This paper is structured as follows. We start by presenting the effective medium theory of space-time metamaterials in Section \ref{sec:effectivemedium}. We derive an analytical expression for the dispersion relation of waves in space-time modulated media, and in Section \ref{sec:FresnelDrag} we discuss the dispersion surfaces in detail, showing how they reveal a Fresnel drag. Next, in Section \ref{sec:effparam}, we map the space-time metamaterial into a static bianisotropic material and determine the effective medium parameters. In Section \ref{sec:MovingMedium} we present an equivalent moving medium and give an interpretation of the Fresnel drag in space-time metamaterials. Finally, we present a transmission line model in Section \ref{sec:transmissionline} and close the paper with conclusions in Section \ref{sec:conclusions}.

\section{Effective medium theory of space-time metamaterials}
\label{sec:effectivemedium}


Maxwell's equations in space-time modulated media,
\begin{eqnarray}
    \nabla\times\mathbf{E} = -\frac{\partial \mathbf{B}}{\partial t}=-\frac{\partial }{\partial t} \left[ \mu(x,t) \mu_0 \mathbf{H}\right], \\
    \nabla\times\mathbf{H} = \frac{\partial \mathbf{D}}{\partial t}= \frac{\partial}{\partial t}  \left[ \epsilon(x,t) \epsilon_0 \mathbf{E}\right],
\end{eqnarray}
can be solved by taking a Bloch-Floquet ansatz \cite{cassedy1963dispersion}, 
\begin{eqnarray}
    \mathbf{E} = \sum_n \mathbf{E}_n e^{i(k+ng)x + ik_y y - i(\omega+n\Omega)t}, \\
    \mathbf{H} = \sum_n \mathbf{H}_n e^{i(k+ng)x + ik_y y - i(\omega+n\Omega)t},
\end{eqnarray}
with wavevectors $k$ and $k_y$ as defined in Fig. \ref{fig:schematics}. Assuming an s-polarized wave without loss of generality, this procedure leads to a system of equations for the $E_z$ and $H_y$ components of the electromagnetic fields. An eigenvalue equation can be written as, 
\begin{eqnarray} \label{eq:eig}
    k\left[ \begin{array}{c} \mathbf{E} \\  \mathbf{H} \end{array} \right] = 
    \left[ \begin{matrix} \mathbf{M}^{EE} && \mathbf{M}^{EH} \\
    \mathbf{M}^{HE} && \mathbf{M}^{HH}  \end{matrix} \right]
    \left[ \begin{array}{c} \mathbf{E} \\  \mathbf{H} \end{array} \right], 
\end{eqnarray}
where $\mathbf{E}$ and $\mathbf{H}$ now stand for column vectors of the Bloch-Floquet amplitudes of $E_z$ and $H_y$, respectively. The block matrices in the above equation depend on $\omega$ and satisfy $\mathbf{M}^{EE} = \mathbf{M}^{HH} $ and their expressions are given in Appendix \ref{sec:AppA}.  

It is useful to rewrite the eigenvalue equations in the basis of forward and backward propagating waves as follows,
\begin{eqnarray} \label{eq:eigP}
    k\left[ \begin{array}{c} \mathbf{E} +\mathbf{H} \\ \mathbf{E} -\mathbf{H} \end{array} \right] = 
      \left[ \begin{matrix} \mathbf{M}^{++} && \mathbf{M}^{+-} \\
    -\mathbf{M}^{+-} && \mathbf{M}^{--}  \end{matrix} \right]\left[ \begin{array}{c} \mathbf{E} +\mathbf{H} \\ \mathbf{E} -\mathbf{H} \end{array} \right].
\end{eqnarray}
Here, 
\begin{eqnarray}
    \mathbf{M}^{++} &=& \mathbf{M}^{EE}+\frac{1}{2}\left( \mathbf{M}^{EH}+\mathbf{M}^{HE}\right), \\ \mathbf{M}^{+-} &=& -\frac{1}{2}\left( \mathbf{M}^{EH}-\mathbf{M}^{HE}\right), \\  \mathbf{M}^{--} &=& \mathbf{M}^{EE}-\frac{1}{2}\left( \mathbf{M}^{EH}+\mathbf{M}^{HE}\right) .
\end{eqnarray}
Solving Eq. (\ref{eq:eigP}) yields the dispersion relation of the system, $k(\omega)$. 

In the long wavelength limit, the dispersion relation can be written analytically by considering only three neighbouring modes in the eigenvalue equation and approximating $\omega\ll\Omega$, $k\ll g$. This gives, 
\begin{equation}
    \label{eq:dispersion}
    \beta^2\omega^2 = \kappa^2k_y^2 + \left(k - \delta\omega \right)^2,
\end{equation}
where
\begin{eqnarray}
     \beta^2 &=& c_m^{-2} \Big(1+\alpha_e^2\frac{2\Omega^2}{c_m^2g^2-\Omega^2} \Big)\Big(1+\alpha_m^2\frac{2\Omega^2}{c_m^2g^2-\Omega^2} \Big) , \\
     \kappa^2 &=& 1+\alpha_m^2\frac{2\Omega^2  }{c_m^2g^2-\Omega^2}, \\
     \delta &=& \alpha_e\alpha_m\frac{2g\Omega }{c_m^2g^2-\Omega^2},
\end{eqnarray}
 and $c_m^2=1/\epsilon_m\mu_m$. 
Hence the dispersion surfaces are approximately circles of radius $\omega\beta$ and whose center is displaced from the origin along the $k$ axis by $\delta\omega$. This produces an asymmetry with respect to the $k=0$ axis, since the dispersion surfaces are displaced along the direction of the modulation, as shown in Fig. \ref{fig:FresnelDrag} and as will be discussed below in detail.

\section{Fresnel drag}
\label{sec:FresnelDrag}

\begin{figure}[t!]
    \begin{center}
        \includegraphics[width=0.95\columnwidth]{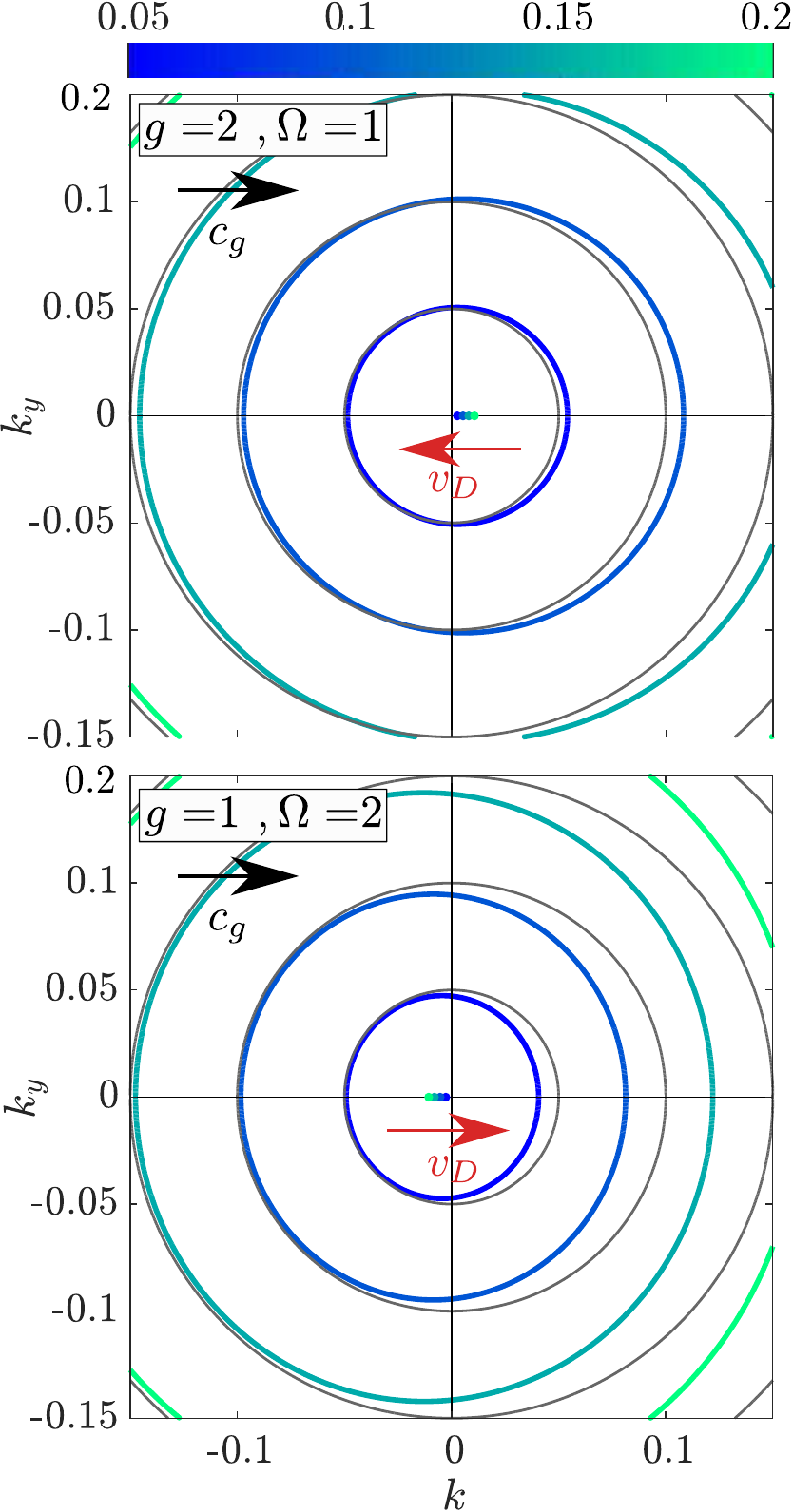}
        \caption{Fresnel drag in space-time $\epsilon$-and-$\mu$ modulated metamaterials with $\alpha_e=\alpha_m=0.2$. The iso-frequency contours in the $(k,k_y)$ plane are shown for (a) subluminal ($g>\Omega$), and (b) superluminal ($\Omega>g$) modulations. Note the change in radius compared to the dispersion in unmodulated media (gray circles), and the displacement along the $k$ axis. In the subluminal case, the space-time modulation drags the waves in the direction opposite to its phase velocity (the effective drift velocity, $v_D$, sketched in red, is anti-parallel to the modulation phase velocity, $c_g$, sketched in black), while the drift velocity flips sign ($v_D$ parallel to $c_g$) upon a superluminal modulation $c_g>c_m$. Note the displacement of the centers of the contours, which are plotted with dots of the same colour as the corresponding contour. The group velocity direction is given by the normal to these dispersion surfaces.}
        \label{fig:FresnelDrag}
    \end{center}
\end{figure}

Here we interpret the displacement of the dispersion surface shown above as an aether drag, which is well known to affect light propagating through moving media \cite{Kong}. In the conventional Fresnel drag effect, moving matter drags light, imparting an extra speed to it. Measured by Fizeau in 1851, Einstein noticed that it is a relativistic effect, and that the speed of light in the medium can be calculated from Lorentz's velocity formula. While it may seem that moving matter is needed to produce a Fresnel drag, our results show that it can also emerge in the presence of space-time modulations, where the modulation profile appears to be moving at a certain speed despite the absence of moving matter. This surprising result can be understood from our effective medium theory. In fact, as we will show in Section \ref{sec:MovingMedium}, the space-time modulated metamaterial can be equivalently represented by a moving medium, which explains the light-dragging effect. Interestingly, the effective drift is present as long as $\delta \neq 0$, i.e. $\alpha_e\cdot\alpha_m \neq 0$ and $g\cdot\Omega \neq 0$. In other words, the Fresnel drag in space-time metamaterials emerges when both the permittivity and permeability are modulated, and when the modulations have non-zero spatial and temporal frequencies.

The isofrequency contours given by Eq. (\ref{eq:dispersion}) are shown in Fig.  \ref{fig:FresnelDrag} for two examples of space-time metamaterials: with subluminal (top) and superluminal (bottom) modulation speeds. The modulation strength is chosen in both cases as $\alpha_e=\alpha_m=0.2$ such that the permittivity and permeability are equally modulated. For comparison, we also show the isofrequency contours of unmodulated media ($\omega^2/c_m^2=k^2+k_y^2$, gray lines). It can be seen how in the presence of the modulation, the circular contours change shape, and their center, which is plotted as a dot of the same color as the corresponding contour, gets displaced along the $k$ axis. This shows the Fresnel drag in space-time metamaterials. Interestingly, when the modulation speed is subluminal, the isofrequency contours are displaced in the direction of the modulation's phase velocity (see top panel). On the other hand, when the modulation is superluminal, the contours are displaced in the direction opposite to the modulation's phase velocity. In order words, given a space-time variation of $\epsilon$ and $\mu$, the direction of the Fresnel drag can be switched by changing between subluminal and superluminal modulations. Remarkably, this implies that the direction of the Fresnel drag velocity is anti-parallel to the phase velocity of the modulation in the subluminal case, and parallel to it for the superluminal case. This seemingly counterintuitive fact will be fully justified below. 

In addition, the extent of the Fresnel effect in these media depends on the magnitude of the modulation speed. It increases as $c_g$ increases from $c_g=0$ in the subluminal regime, and as $c_g$ decreases from $c_g\rightarrow\infty$ in the superluminal regime. As the luminal modulation regime $c_g\to c_m$ is approached, however, strong interaction between multiple bands results in an unstable regime dominated by intraband photonic transitions~\cite{cassedy1963dispersion,galiffibroadbandnonreciprocal2019}, which spoils the separation of length scales needed for long-wavelength homogenisation.

\section{Effective bianisotropic parameters and non-reciprocity}
\label{sec:effparam}
Next we derive the effective medium parameters of the space-time metamaterial. We start by noting that the dispersion relation of waves in the space-time modulated medium, Eq. (\ref{eq:dispersion}), is of the form of the dispersion of waves in a bianisotropic medium with uniaxial $\epsilon$ and $\mu$,
\begin{eqnarray}
    \boldsymbol{\epsilon} = \left[ 
    \begin{array}{ccc} 
        \epsilon_x & 0 & 0 \\
        0 & \epsilon & 0 \\
        0 & 0 & \epsilon 
    \end{array} \right] ; \;
    \boldsymbol{\mu} = \left[ 
    \begin{array}{ccc} 
        \mu_x & 0 & 0 \\
        0 & \mu & 0 \\
        0 & 0 & \mu 
    \end{array} \right],
\end{eqnarray}
and magnetoelectric coupling,
\begin{eqnarray}
    \boldsymbol{\xi} = \boldsymbol{\zeta}^T =\left[ 
    \begin{array}{ccc} 
        0 & 0 & 0 \\
        0 & 0 & +\xi \\
        0 & -\xi & 0 
    \end{array} \right]. 
\end{eqnarray}
For s-polarized waves we have in this case,
\begin{equation}
    \mu\mu_0\epsilon\epsilon_0\omega^2 = \frac{\mu}{\mu_x}k_y^2 + \left(k - \xi\omega \right)^2.
\end{equation}
By comparing to Eq. (\ref{eq:dispersion}) we can identify a set of effective medium parameters for the space-time modulated metamaterial, 
\begin{eqnarray}
    \epsilon_x &=& \mu_x = 1 \label{eq:effparam1}\\
    \epsilon &=&  1 + \alpha_e^2\frac{2\Omega^2}{c_m^2g^2-\Omega^2} \\
    \mu &=&  1 + \alpha_m^2\frac{2\Omega^2}{c_m^2g^2-\Omega^2} \\
    \xi &=& \alpha_e\alpha_m\frac{2g\Omega}{c_m^2g^2-\Omega^2}.  \label{eq:effparam2}
\end{eqnarray}
This shows that the modulated medium can be represented by an effective bianisotropic material. Interestingly, the magnetoelectric coupling vanishes if the modulation is only spatial ($\Omega=0$) or only temporal ($g=0$), or if only one of the electromagnetic parameters is modulated, i.e. $\alpha_{e}\cdot \alpha_{m}=0$. We also note in passing that space-time metamaterials with only $\epsilon$ or $\mu$ modulations are mapped to uniaxial media (without any magnetoelectric coupling).

\begin{figure}[t!]
    \begin{center}
        \includegraphics[width=\columnwidth]{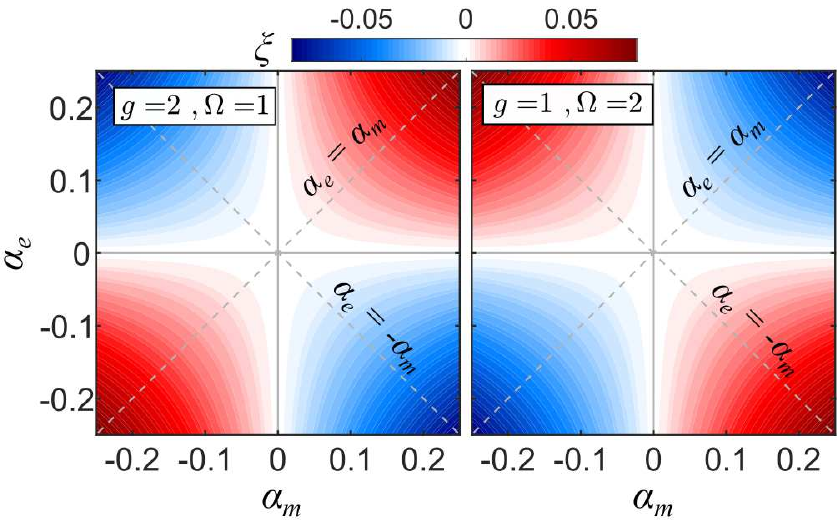}
        \caption{Non-reciprocity map in the $\alpha_e,\alpha_m$ parameter space for subluminal (left) and superluminal gratings (right). The colour map shows effective magnetoelectric coupling, $\xi$, which is equal to $\delta $ and which also gives the difference in the inverse of phase velocities between forward and backward propagating waves, $\Delta=(k_+ - |k_-|)/\omega = 2 \delta $. It is seen that it is necessary to modulate both parameters in order to achieve non-reciprocity in the long wavelength limit.}
        \label{fig:Delta}
    \end{center}
\end{figure}

The magnitude of $\xi$ is presented in Fig. \ref{fig:Delta}, as a function of the electric and magnetic modulation strengths. When changing from a subluminal (left) to a superluminal (right) modulation speed, the magnetoelectric coupling changes sign, in agreement with the reverse directions obtained for the Fresnel drag. In addition, we can conclude from this plot that a phase difference of $\pi$ between the $\epsilon$ and $\mu$ modulations also reverses the direction of the aether drag with respect to in-phase modulations, since $\xi$ changes sign when $\text{sign}(\alpha_e\cdot\alpha_m)<0$.  

In agreement with the effective bianisotropic parameters, the dispersion surfaces shown in Fig.~\ref{fig:FresnelDrag} represent non-reciprocal dispersion relations, a result of the time-reversal symmetry breaking induced by the modulation. From Eq. (\ref{eq:dispersion}) we can explicitly write dispersion relations for forward and backward propagating waves, 
\begin{equation} 
    \label{eq:dispersionpm}
    k_{\pm} = \delta \omega \pm \sqrt{ \beta^2 \omega^2 - \kappa^2 k_y^2 }
\end{equation}
Clearly, when $\delta\neq0$ ( $\xi\neq0$) the two branches are asymmetric with respect to $k=0$, that is, the system is non-reciprocal when both $\alpha_e$ and $\alpha_m$ are non-zero. For waves travelling in the direction of the modulation, $k_y=0$, it is easy to see that the difference between the wavevectors of forward and backward waves at a given frequency is given by $\Delta=2\delta=2\xi$. Hence, we can interpret Fig. \ref{fig:Delta} as a map of the strength of non-reciprocity. From this we confirm that a non-reciprocal response requires both permittivity and permeability modulations, and that for a fixed total modulation strength, $\alpha_e^2+\alpha_m^2=\alpha^2$, non-reciprocity is maximized by $\alpha_e=\alpha_m$.

\begin{figure*}[ht!]
    \centering
    \includegraphics[width=0.9\textwidth]{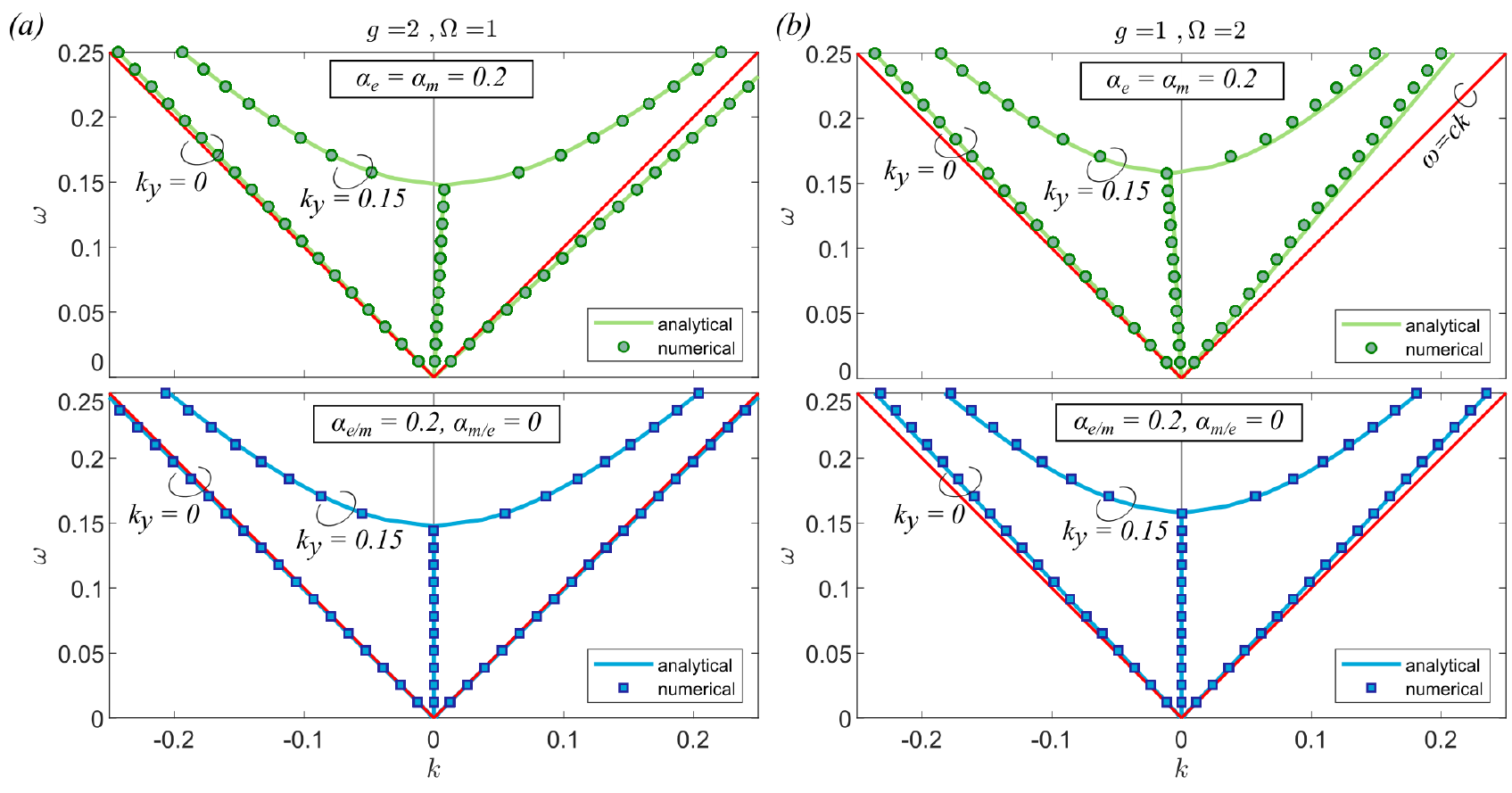}
    \caption{Dispersion relations for $k_y=0$ and $k_y=0.15$, for subluminal (a) and superluminal (b) metamaterials. The top panels show how $\epsilon$ and $\mu$ travelling-wave modulations yield non-reciprocal dispersion. The permittivity and permeability modulations are set equal, $\alpha_e=\alpha_m=0.2$. The bottom panels show how when only one of the parameters is modulated with the same strength as before ($\alpha_e=0.2$ or $\alpha_m=0.2$) and the other is kept constant ($\alpha_m=0$ or $\alpha_e=0$), the system is reciprocal. The dots represent numerical results and the dashed lines show the effective medium approximation. The red lines represent the light lines}
    \label{fig:bands}
\end{figure*}

\section{Equivalent moving medium}
\label{sec:MovingMedium}
The link between moving media and bianisotropic effective parameters has been shown in the past~\cite{vehmas2014transmission,Silveirinha2014spontaneous}. Hence, since the space-time modulated medium can be represented by effective bianisotropic parameters, it can also be linked to an equivalent moving medium. Below we derive the speed of the equivalent moving medium.

 Consider a moving uniaxial medium with parameters, $\hat{\boldsymbol{\epsilon}}' = \text{diag}(\epsilon'_x,\epsilon',\epsilon') $, $\hat{\boldsymbol{\mu}}' = \text{diag}(\mu'_x,\mu',\mu') $, moving along the $x$ axis with speed $v$. Lorentz transformations tell us that in the moving frame the permittivity and permeability change as \cite{Kong},
 \begin{eqnarray} 
    \label{eq:movingparameters}
    \epsilon_x=\epsilon'_x, \, 
    \epsilon = \epsilon'\frac{1-v^2/c_m^2}{1-\epsilon'\mu'v^2/c_m^2} \\
    \mu_x=\mu'_x, \, 
    \mu = \mu'\frac{1-v^2/c_m^2}{1-\epsilon'\mu'v^2/c_m^2} .
 \end{eqnarray}
Also, the electric and magnetic fields in the moving frame are coupled through magnetoelectric tensors, 
\begin{eqnarray}
\label{eq:movingparameters2}
    \boldsymbol{\xi} = \boldsymbol{\zeta}^T =\left[ 
    \begin{array}{ccc} 
        0 & 0 & 0 \\
        0 & 0 & +\xi \\
        0 & -\xi & 0 
    \end{array} \right], 
\end{eqnarray}
where 
\begin{align}
    \xi = \frac{v}{c_m} \frac{\epsilon'\mu'-1}{\epsilon'\mu'-v^2/c_m^2}.
\end{align}

By mapping the tensors (\ref{eq:movingparameters}-\ref{eq:movingparameters2}) to the set of effective bianisotropic parameters in Eqs. (\ref{eq:effparam1}-\ref{eq:effparam2}), we determine,
\begin{align} \label{eq:effvelocity}
    v_D \approx -\xi c_m^2 \epsilon'\mu'\approx - \alpha^2\frac{2c_m^2g\Omega}{c_m^2g^2-\Omega^2} ,
\end{align}
and
\begin{align} \label{eq:movingparam}
        \epsilon'=\mu'\approx \frac{\epsilon}{1-v^2/c_m^2}  \approx  1 + \alpha^2\frac{2\Omega^2}{c_m^2g^2-\Omega^2} ,
\end{align}
where we have restricted ourselves to the case where $\epsilon=\mu$, i.e., $\alpha_e=\alpha_m$, for simplicity. The velocity $v_D$ is the drag velocity that light experiences in the space-time metamaterial. This justifies our interpretation of the isofrequency contours in Fig. \ref{fig:FresnelDrag} as an aether drag.  

It is clear from Eq. (\ref{eq:effvelocity}) that when the modulation is subluminal ($g>\Omega$), $v_D<0$, such that the effective medium moves in the opposite direction to the phase velocity of the modulation. Hence, forward propagating modes slow down in the presence of subluminal space-time modulations, and backward propagating modes speed up. On the other hand, for superluminal modulations ($g<\Omega$), $v_D>0$, such that the drag velocity points in the same direction as the modulation's phase velocity. In this case, forward waves speed up and backward waves slow down. We remark here that there are two effects at work in the space-time modulated metamaterial. In addition to the Fresnel drag, there is an increase/reduction in the effective permittivity and permeability for subluminal/superluminal media, as can be seen in Eq. (\ref{eq:movingparam}).
Hence, the slowing down or speeding up of the waves is in relation to waves propagating in the medium with reduced/increased effective $\epsilon$ or $\mu$, rather than with respect to $c_m$.

Importantly, relativity imposes that the permittivity and permeability of a moving medium change at the same pace, see  (\ref{eq:movingparameters}-\ref{eq:movingparameters2}), as electric and magnetic effects are intimately linked to each other in relativity. Hence, the mapping to a moving medium can be done only in the case of space-time modulations of both $\epsilon$ and $\mu$, whereas the moving medium interpretation is not exact if either material parameter is not modulated. This is a fundamental difference between systems where one between electric and magnetic properties is modulated in space and time, and systems which feature both modulations simultaneously, and explains why the aether drag and non-reciprocity disappear for modulations of only $\epsilon$ or $\mu$.
 
\section{Non-reciprocal dispersion relations}
\label{sec:dispersions}
We now discuss in detail the effective medium theory presented in Section \ref{sec:effectivemedium} and validate our analytical expressions against numerical results. 

The non-reciprocal dispersion relations given by Eq. (\ref{eq:dispersionpm}) are shown in Fig. \ref{fig:bands} for a choice of non-zero electric and magnetic modulation amplitudes $\alpha_e=\alpha_m=0.2$ (top panels, solid green lines). Numerical results obtained from Eq. (\ref{eq:eig}) are also plotted with dots,  validating our effective medium theory. For waves propagating in the direction of the modulation ($k_y=0$), the dispersion relations are linear, $k_{\pm} = (\delta \pm  \beta) \omega $, as can be seen in the plot. In fact, when  $\alpha_e=\alpha_m$, the eigenvalue equation \ref{eq:eigP} becomes block-diagonal because $\boldsymbol{M}^{EH}=\boldsymbol{M}^{HE}$ and $\boldsymbol{M}^{+-}=\boldsymbol{M}^{-+}=0$, reflecting the fact that forward and backward waves do not interact in this case. Indeed this periodic system, being impedance-matched to free space at all positions $x$ and times $t$, has zero band gaps at higher frequencies and momenta, as discussed in Ref. \cite{Taravati2018}. However, despite the fact that forward and backward waves do not interact with each other, the modulation acts on each of them, yielding a non-reciprocal response. The speed of forward and backward waves is in fact different, and, according to Eq. (\ref{eq:dispersionpm}), it is given respectively by
\begin{eqnarray}
    v_+ &=& \frac{\omega}{k_+} = c_m\left(1+2\alpha^2\frac{c_g}{c_m-c_g}\right)^{-1}, \\
    v_- &=& \frac{\omega}{k_-} = -c_m\left(1-2\alpha^2\frac{c_g}{c_m+c_g}\right)^{-1}.
\end{eqnarray}

It is also clear from the above equations that for subluminal grating speeds ($c_g<c_m$), forward waves propagating through the space-time modulated medium speed up and backward waves slow down, in agreement with a negative Fresnel drag velocity. In particular, $v_+<c_m$ and $|v_-|>c_m$, such that the branch at $k>0$ lies below the light line for the unmodulated material, and the branch at $k<0$ lies above it, as can be seen in the top panel (a). For the case of superluminal grating speeds ($c_g>c_m$), the speed of forward waves increases and that of backward waves decreases, in agreement with a positive Fresnel drag velocity. We remark here that while both velocities increase with respect to waves propagating in the absence of modulation ($v_+>c_m$, $|v_-|>c_m$, both branches lie above the light line), the Fresnel drag in fact acts with respect to a medium with reduced effective permittivity and permeability (as given by Eq. (\ref{eq:movingparam})). In the superluminal case the phase velocity of waves in such medium is in fact reduced by a large amount, $\sim \Omega^2$, and the reciprocity breaking term is not strong enough to move the backward propagating branch to the opposite side of the light line. Hence, while the drag acts in the same direction for both sets of waves, both branches lie above the light line of unmodulated media [top panel, (b)]. These different phase velocities for forward and backward modes are consistent with the results presented in Fig. \ref{fig:Delta}, where $\Delta/2=(v_g^+)^{-1}-(v_g^-)^{-1} =\delta$ is plotted. 

On the other hand, if either the permittivity or the permeability are kept constant ($\alpha_{e/m}=0.2$, $\alpha_{m/e}=0$), the reciprocity-breaking term $\delta=0$, and the dispersion relations are fully reciprocal in the long wavelength limit, as shown in the lower panels. In this case, and for subluminal modulations, the forward/backward wave decreases/increases in speed with respect to $c_m$, and the dispersion relations lie below the light lines, and conversely for superluminal modulations, for which both branches lie above the light lines. The phase velocities are given by $\beta^{-1}$ in this case, whose deviation from the unmodulated system is quantified by terms $\sim \Omega^2/(c_m^2g^2-\Omega^2)$. This explains why the effect is more pronounced for superluminal (b) than subluminal modulations (a), and why for superluminal both branches remain above the light line with electric and magnetic modulations.  

Finally, for off-normal incidence, $k_y\neq0$, the dispersion presents a cut-off and exhibits the Fresnel drag by tilting towards the positive $k$ direction for subluminal modulations and towards the negative one for superluminal modulations, in agreement with Fig.~\ref{fig:FresnelDrag}.

\section{Transmission line model}
\label{sec:transmissionline}

We propose a realization of Fresnel drag in a transmission line comprising the discrete elements shown in Fig~\ref{fig:transmission_line}a, which can be modelled as detailed in Appendix \ref{sec:AppB}. Varactors are used, so that a strong pump signal can modulate their capacitance in time, with appropriate phasing from one varactor to another. Varactors are available with a typical capacitance of 100 pF. Both the inductive and capacitive elements must be modulated in order to see the drag effect: ferrite-core inductors, with inductance of the order of 1$\mu$H are commercially available. The ferrite core has a non-linear response, enabling the pump beam to vary the inductance. A powerful high frequency pump signal is sent through the system to produce the required modulation of the elements. Simultaneously, a low frequency probe tests the effect of the modulation, as illustrated in Fig.~\ref{fig:transmission_line}b. For the values of inductance and capacitance quoted, the cut-off frequency of the transmission line would be approximately $6$ MHz.

One suggestion for detecting the drag is to make an analogy with the electrons confined to a loop. In the absence of a magnetic field, non zero angular momentum states are twofold degenerate, as dictated by time reversal symmetry. A magnetic field breaks this symmetry and splits the formerly degenerate states. Fig.~\ref{fig:transmission_line}c shows the configuration. The loop quantizes the states, and the space-time modulation of the elements produces an effective gauge field for photons~\cite{fang2012realizing}, resulting in a splitting of opposite-propagating modes, in analogy with the Zeeman splitting in atomic physics~\cite{fleury2014sound, estep2014magnetic}.

\begin{figure}
    \centering
    \includegraphics[width=\columnwidth]{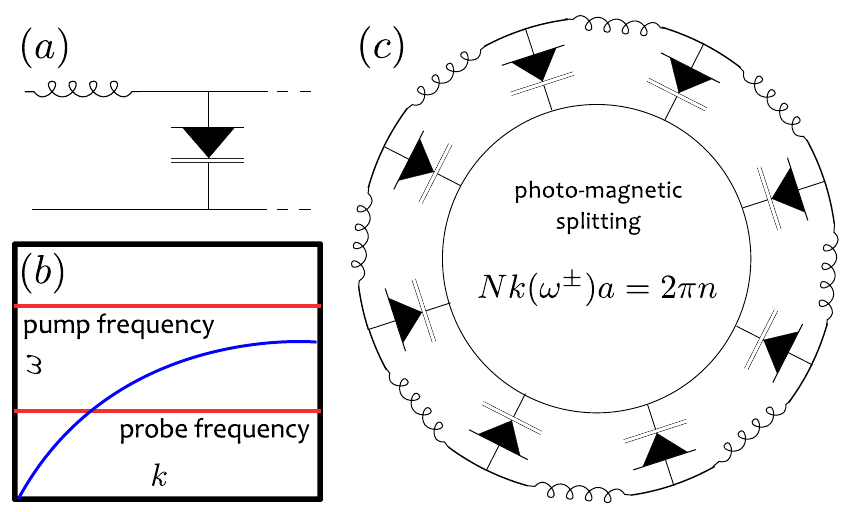}
    \caption{(a) The capacitor in a conventional transmission line is replaced by a varactor, whose capacitance varies with the voltage across it. The inductors are also assumed to vary with the bias current. (b): Schematic dispersion showing the pump and probe frequencies. (c) Discrete elements combined into a loop to demonstrate the presence of a photo-magnetic field.}
    \label{fig:transmission_line}
\end{figure}

\section{Conclusions}
\label{sec:conclusions}
Space-time modulated systems can, in the low frequency limit, be represented as bianisotropic metamaterials, or alternatively as uniaxial metamaterials in motion relative to the observer's reference frame. In the latter case we claim that this is an instance of Fresnel's aether drag hypothesis but with the interesting twist that modulations of the system do not displace the physical medium itself and the motion is apparent rather than real. Further the velocity of the modulations, $c_g$, is only indirectly related to the drag velocity, $v_D$, whose sign and magnitude can be changed by the size and relative phase of the electric and magnetic modulation amplitudes. Our system presents a rich variety of phenomena with parameters that are highly tuneable.

We mention in passing that we expect realisations in other wave phenomena, such as acoustics, where both density and bulk modulus need to be modulated to achieve a Willis stress-velocity coupling parameter in the homogenisation regime \cite{nassar2017modulated}, to be possible. Finally, we suggest an electronic circuit model as a possible test bed for these ideas.

\appendix
\section{}
\label{sec:AppA}
The matrices in the eigenvalue equation, (\ref{eq:eig}), are given by: 
 \begin{eqnarray}
     \mathbf{M}^{EE}_{n'n} &=& \mathbf{M}^{HH}_{n'n} = -ng\delta_{nn'}, \\
     \mathbf{M}^{HE}_{n'n} &=& -\epsilon_0  
         \begin{Bmatrix} 
             (\omega+n\Omega)\delta_{nn'}-k_y^2c_m^2\left(S^{EH}_{n'n} \right)^{-1} \\ 
             + \alpha_e (\omega+n\Omega)  (\delta_{n+1,n'}+\delta_{n-1,n'})  \\ 
             + \alpha_e \Omega (\delta_{n+1,n'}-\delta_{n-1,n'})
          \end{Bmatrix},     \label{eq:Mhe}\\
     \mathbf{M}^{EH}_{n'n} &=& -\mu_0  
         \begin{Bmatrix} 
             (\omega+n\Omega)\delta_{nn'} \\ 
             + \alpha_m (\omega+n\Omega)  (\delta_{n+1,n'}+\delta_{n-1,n'})  \\ 
             + \alpha_m \Omega (\delta_{n+1,n'}-\delta_{n-1,n'})
          \end{Bmatrix} ,
 \end{eqnarray}
 with 
 \begin{eqnarray}
     \mathbf{S}^{EH}_{n'n} &=&   
         \begin{Bmatrix} 
             (\omega+n\Omega)\delta_{nn'}  \\ 
             + \alpha_m (\omega+n\Omega)  (\delta_{n+1,n'}+\delta_{n-1,n'})  \\ 
             + \alpha_m \Omega (\delta_{n+1,n'}-\delta_{n-1,n'})
          \end{Bmatrix} .
 \end{eqnarray}

\section{}
\label{sec:AppB}
The elements in the transmission line in Fig. 5 vary in time as
\begin{eqnarray}
    C_n(t) = B_1\left[  1 + \alpha_1\left( e^{+i(\Omega t+\theta_n)} + e^{-i(\Omega t+\theta_n)} \right) \right]^{-1} \\
    L_n(t) = B_2\left[  1 + \alpha_2\left( e^{+i(\Omega t+\theta_n)} + e^{-i(\Omega t+\theta_n)} \right) \right]^{-1}
\end{eqnarray}
Solving the equations, 
\begin{eqnarray}
    \dot{V}_n = C_n^{-1} (I_n-I_{n+1}) \\
    \dot{I}_{n+1} = L_{n+1}^{-1} (V_n-V_{n+1})
\end{eqnarray}
it can be proven that our transmission line model reproduces the Fresnel drag when both elements are modulated.

\begin{acknowledgments}
P.A.H. acknowledges funding from Funda\c{c}\~ao para a Ci\^encia e a Tecnologia and Instituto de Telecomunica\c c\~oes under project CEECIND/03866/2017. P.A.H and R.V.C. acknowledge funding from the Leverhulme Trust. E.G. acknowledges support through a studentship in the Centre for Doctoral Training on Theory and Simulation of Materials at Imperial College London funded by the EPSRC (EP/L015579/1). J.B.P. acknowledges funding from the Gordon and Betty Moore Foundation. S.G. and R.V.C. acknowledge support from the EPSRC Programme Grant (EP/L024926/1).
\end{acknowledgments}

\bibliographystyle{unsrt} 
\bibliography{apssamp}

\end{document}